\documentclass[twocolumn,showpacs,preprintnumbers,amsmath,amssymb]{revtex4}
\usepackage{dcolumn}
\usepackage{bm}
\usepackage{graphicx}

\begin{document}
\title{Wave-packet trains of a time-dependent harmonic oscillator}
\author{Wenhua Hai$^{1,2}$ \ \ Shengxun Huang$^1$ \ \ Kelin Gao$^2$ }
\affiliation{ $^1$Department of physics, Hunan Normal University,
Changsha
410081, China \\
$^2$Laboratory of Magnetic Resonance and Atomic and Molecular
Physics, \\ Wuhan Institute of Physics and Mathematics, Chinese
Academy of Sciences,\\ Wuhan 430071, China}

\altaffiliation{Correspondence Author: W. Hai.}

\email{adcve@public.cs.hn.cn}
\begin{abstract}
By using a test-function method, we construct $n$ exact solutions
of a quantum harmonic oscillator with a time-dependent "spring
constant". Any $n$-th solution describes a wave-packet train
consisting of $n+1$ packets. Its center oscillates like the
classically harmonic oscillator with variable frequency, and width
and highness of each packet change simultaneously. When the
deformation is small, it behaves like a soliton train, and the
large deformation is identified with collapse and revival of the
wave-packet train.

\end{abstract}

\pacs{03.65.Ge, 42.50.Gy, 32.80.Pj, 42.50.Vk}

\maketitle

Classical and quantum motions of the time-dependent harmonic
oscillator are practically important, because of their physical
realization in a Paul trap \cite{Paul}, \cite{Dehmelt}. For some
periodic and time-dependent "spring constants", the classical
motions are governed by the well-known Mathieu equation
\cite{Mclachlan}. Most of works for the system are of classical
dynamics, such as the stability analysis \cite{Dawson}, order and
chaos \cite{Brewer}, \cite{Blumel}, and so on. In the ultracold
ion case, quantum motion becomes important for practical
applications \cite{Diedrich}, \cite{Wu}. For convenience' sake,
many works based on quantum mechanics, say, the quantum
computations \cite{Poyatos}, \cite{Pachos} and state preparations
\cite{Monroe}, \cite{Zheng}, were performed by adopting the
pseudopotential model \cite{Dehmelt2} with constant trapped
frequencies. There exist a few of works on the quantum motions in
a Paul trap with periodically variable trapped frequencies. Cook
and coworkers proximately treated the time-dependent quantum
oscillator by perturbing about a harmonic oscillator solution for
a time-independent, effective potential \cite{Cook}. Combescure,
Brown and Feng \cite{Combescure}-\cite{Feng} gave similar exact
solutions of the time-dependent system by employing different
methods respectively. Their results showed that quantum-mechanical
solutions of the system depend on the corresponding classical
ones.

Very recently, Moya-Cessa and Guasti reported a coherent state
solution of the time-dependent quantum harmonic oscillator
\cite{Cessa} that may be associated with summing Brown's $n$ exact
solutions. For a time-independent trapped frequency, we suggest a
test-function method to get $n$ exact solutions of the system
\cite{Hai}. In the present paper, we shall adopt this method to
find $n$ exact solutions of the time-dependent quantum harmonic
oscillator. These solutions describe $n$ wave-packet trains that
propagate and breathe simultaneously. By selecting initial
conditions to fix centers of the trains, the result is reduced to
Brown's one \cite{Brown}. And the ground state with $n=0$ may be
related to Cessa's coherent state \cite{Cessa}.

We consider a single ion confined in a Paul trap that provides an
oscillating quadrupole potential with the resulting quantum motion
governed by the time-dependent Schr$\ddot o$dinger equation
\begin{eqnarray}
i \frac{\partial \psi}{\partial t}=- \frac {1}{2} \frac{d^2
\psi}{dx^2} + \frac {1}{2}k(t) x^2 \psi,
\end{eqnarray}
where the "spring constant" is a periodic function of time,
$k(t)=U^2 + V \cos \omega t$ with $\omega$ being the rf-driven
frequency and $U^2, V$ the trap parameters. We have adopted the
unit with $m=\hbar =1$ and will normalize time in unit $2/ \omega$
so that $k(t)=U^2 + V \cos 2t$. The spatial coordinate $x$ and
probability density $|\psi|^2$ are normalized by the harmonic
oscillator length $l_h=\sqrt{\hbar /(m \omega)}$ and inverse one
$l_h^{-1}$, respectively. Consider the test function as a solution
of Eq. (1) in the form
\begin{eqnarray}
\psi_n=a_n(t)H_n(\xi)\exp [b(t)x-c(t)x^2-f^2(t)/2], \ \nonumber \\
\xi=e(t)x-f(t). \ \ \ \ \ \ \ \ \ \ \ \ \ \ \ \ \ \
\end{eqnarray}
Here $a_n(t), \ b(t), \ c(t)$ are the complex functions of time
and $e(t), \ f(t)$ the real functions. Applying Eq. (2) to Eq.
(1), we arrive at the equation
\begin{eqnarray}
e^2 \frac{\partial^2 H_n}{\partial \xi^2}&+&2(be-i\dot f+i\dot e
x-2cex)\frac{\partial H_n}{\partial \xi} \nonumber
\\ + 2\Big[i\frac{\dot {a}_n}{a_n}&-& i f\dot f+\frac{b^2}{2}-c
+(i\dot b-2bc)x \nonumber \\ &+& \Big(2c^2-i\dot c-\frac 1 2 k(t)
\Big)x^2\Big]H_n=0.
\end{eqnarray}
Noticing the Hermitian equation $\partial^2 H_n /
\partial \xi^2 -2\xi \partial H_n / \partial \xi +2nH_n =0$, Eq.
(3) implies
\begin{eqnarray}
i \dot c =2c^2-k(t)/2, \ \ i \dot b =2bc, \ \nonumber \\ i \dot e
=2ce-e^3, \ \ i \dot f =be- e^2f, \ \ \nonumber \\   i \dot {a}_n
/a_n =i f\dot f -b^2/2+c+ne^2.
\end{eqnarray}

The first of Eq. (4) is a complex Riccati equation, which can be
changed into a complex equation of a classical harmonic oscillator
\begin{eqnarray}
\ddot \varphi =-k(t) \varphi = -(U^2 + V \cos 2t)\varphi,
\end{eqnarray}
through the function transformation $c=\dot \varphi /(2i
\varphi)$. This equation is just the Mathieu's one whose exact
solution can be taken as the form of infinite trigonometrical
series \cite{Mclachlan}. Generally, for given physical parameters
$U^2, V, \omega$ and initial conditions to determine coefficients
of the infinite series is difficult. In order to overcome this
difficulty, we here shall employ an iteration method to seek the
series solution, through an integration equation. Setting real
part and imaginary part of $\varphi$ as $\varphi_1$ and
$\varphi_2$ respectively, they satisfy same equation (5). The
corresponding integration equations read as  \cite{Hai2}
\begin{widetext}
\begin{eqnarray}
\varphi_1&=&A\cos(U t+\alpha)+VU^{-1}\Big(\sin Ut \int_{0}^{t}
\cos Ut \cos 2t \varphi_1 dt -\cos Ut\int_{0}^{t} \sin Ut \cos 2t
\varphi_1 dt \Big), \nonumber \\ \varphi_2&=&B \cos(U t+\beta)
+VU^{-1}\Big(\sin Ut \int_{0}^{t} \cos Ut \cos 2t \varphi_2dt-\cos
Ut\int_{0}^{t} \sin Ut \cos 2t \varphi_2dt \Big),
\end{eqnarray}
\end{widetext}
where $A, \ B, \ \alpha$ and $\beta$ are some real constants
associated with initial conditions of the classical oscillator. In
quantum-mechanical treatment, there exist many possibilities on
these constants that correspond to different states $\psi$. The
two equations in Eq. (6) can be directly proved by inserting them
into Eq. (5). Applying Eq. (6), we construct the solution of Eq.
(5) as
\begin{eqnarray}
\varphi &=& \varphi_1+i \varphi_2=\rho(t)e^{i\theta(t)}, \nonumber
\\ \rho(t)&=& \sqrt{\varphi_1^2+\varphi_2^2}, \ \ \ \theta(t)=\arctan
\frac{\varphi_2}{\varphi_1}.
\end{eqnarray}
The real functions $\rho (t)$ and $\theta (t)$ will be use as the
known functions in follows. So far, most of experimental works and
theoretical analyses on the Paul-trapped ions are focused on the
first stability region \cite{Blumel2}, where the trap parameters
$U^2$ and $V$ are small, $U^2<1, \ V<1$ and $V \sim U^2<U$. In
such case, we can treat the terms proportional to $V$ of Eq. (6)
as perturbations and adopt the iteration method to produce the
perturbed solution.

Returning to the transformation between $\varphi$ and $c$ yields
\begin{eqnarray}
c=\frac{\dot \varphi}{2i \varphi}= \frac 1 2 \dot \theta -i
\frac{\dot \rho}{2\rho}.
\end{eqnarray}
Substitution of Eq. (7) into Eq. (5) yields equations of the phase
$\theta$ and module $\rho$ as
\begin{eqnarray}
\ddot \theta =-2\dot \theta \dot \rho / \rho, \ \ddot \rho=\rho
\dot \theta ^2 -k(t)\rho
\end{eqnarray}
with the first integration
\begin{eqnarray}
c_0=\rho^2 \dot \theta =\varphi_1 \dot \varphi_2-\varphi_2 \dot
\varphi_1.
\end{eqnarray}
Combining Eqs. (8) with Eq. (4) and applying the relation (10), we
easily obtain
\begin{eqnarray}
b&=&b_0 \frac{\exp(-i\theta)}{\rho}, \ e=\frac{\sqrt{c_0}}{\rho}
=\sqrt{\dot \theta},    f=\frac{b_0}{\sqrt{c_0}}\cos \theta ,
\nonumber \\ a_n&=&\frac{A_0}{\sqrt{\rho}}\exp
\Big\{-i\Big[\Big(\frac 1 2 +n\Big)\theta -\frac{b_0^2}{4c_0}\sin
2\theta\Big]\Big\},
\end{eqnarray}
where $b_0$ and $A_0$ are arbitrary constants. Inserting these
into Eq. (2) leads to the exact solution $\psi_n(x, t)$, and the
normalization condition $\int |\psi_n|^2
dx=A_0^2\sqrt{c_0^{-1}}\int
H_n^2(\xi)\exp(-\xi^2)d\xi=A_0^2\sqrt{\pi c_0^{-1}}2^n n!=1$ gives
the constant $A_0=[\sqrt{c_0}/(\sqrt{\pi}2^n n!)]^{1/2}$ such that
from Eqs. (2), (8) and (11) we have the normalized wave-function
\begin{widetext}
\begin{eqnarray}
\psi_n &=& R_n(x, t) \exp[i \Theta _n(x, t)], \ \ n=0, \ 1, \ 2, \cdots, \nonumber \\
R_n &=& \Big[\frac{\sqrt{c_0}}{\sqrt{\pi} 2^n n! \rho
(t)}\Big]^{1/2}H_n(\xi) \exp \Big(- \frac {1}{2}\xi^2 \Big), \
\ \xi=\frac{\sqrt{c_0}x}{\rho (t)}-\frac{b_0}{\sqrt{c_0}}\cos \theta(t), \ \ \ \\
\Theta _n&=& \frac{\dot \rho (t) x^2}{2\rho (t)}-\frac{b_0 x}{\rho
(t)}\sin \theta(t) +\frac{b_0^2}{4c_0}\sin [2\theta(t)]
-\Big(\frac 1 2 +n \Big)\theta(t). \nonumber
\end{eqnarray}
\end{widetext}

The module $R_n$ of exact solution (12) for different quantum
number $n$ describes the wave-packet trains consisting of $n+1$
packets. Orbit of center of the wave-packet trains $x_c(t)$ is
given by $\xi =0$. Applying Eq. (7), from $\xi =0$ we have the
orbit
\begin{eqnarray}
x_c =\frac{b_0}{c_0}\rho (t) \cos \theta(t) = \frac{b_0}{c_0}
\varphi_1,
\end{eqnarray}
which is proportional to real part of the complex solution (7),
namely the orbit of a time-dependent classical oscillator.
Comparing our Eq. (12) with Eq. (24) of Brown's paper
\cite{Brown}, we find that if one let the arbitrary constant $b_0$
be zero, our solution agrees with Brown one. The orbit equation
(13) indicates difference between behaviors of the two solution:
our wave-packet trains oscillate their centers, but the centers of
Brown's wave-packets are rested.

The function $\rho (t)$ appearing in $\xi$ describes widths of the
wave-packet trains and each packet. For any train the average
width of the packets is $\rho(t)/\sqrt{c_0}$. Same function
appearing in radical of Eq. (12) governs highnesses of every
packets. So we call $\rho(t)$ the function of width and highness.
When the changes of the widths and highnesses are small, behavior
of the wave-packets seems to be that of the soliton trains. The
greatly variable widths and highnesses show collapse and revival
of the wave-packet trains, like behaviors of multiple breathers.
The normalization condition implies that the broader wave-packet
train is associated with smaller mean highness and the narrower
wave-packet train corresponds to larger mean highness.

Phase of the exact solution (12) contains the term $(1 /2 +n)
\theta(t)$ that infers average energy $E_n(t)=\langle \psi_n|i
\partial / \partial t|\psi \rangle$ to be proportional to $(1 /2 +n)$. In the
processes of motions, the wave-packet trains may spontaneously
transit from states of higher average energies to that of lower
ones. Some perturbations also could cause transitions between the
states of different quantum numbers. The ground state $\psi_0$ of
Eq. (12) may be similar to Cessa's coherent state \cite{Cessa}. We
shall numerically illustrate the analytical results for the system
parameters $U=0.5, \ V=0.05$ as follows.

\bf The solitonlike trains \rm To show the solitonlike behavior,
we require larger amplitude of Eq. (13) and keep small change of
module $\rho$ such that the wave-packet trains described by Eq.
(12) can propagate through greater distance with small
deformation. For simplicity, we take the parameter set $\alpha=0,
\ \beta=-\pi/2, \ A=B=c_0=1, \ b_0=-10$ and make iteration from
Eq. (6) only to the first order. That is, let $\varphi_1^{(0)}$
and $\varphi_2^{(0)}$ be $\cos Ut$ and $\sin Ut$ respectively, and
use them instead of $\varphi_1$ and $\varphi_2$ in the
corresponding integrations of Eq. (6), obtaining the approximate
solutions
\begin{eqnarray}
\varphi_1 &\approx& \cos 0.5t+0.05 \sin 0.5t \sin 2t \nonumber \\
&+& 0.025
\cos 0.5t \ (1-\cos 2t), \nonumber \\
\varphi_2 &\approx& \sin 0.5t+0.025\sin 0.5t \ (1-\cos 2t).
\end{eqnarray}
Applying the first of Eq. (14) to Eq. (13) yields the orbit of
wave-packets' center. Obviously, the center of wave-packet train
oscillates in the spatial region $-10\le x_c \le 10$ with period
$4\pi$. Combining Eq. (14) with Eq. (7), we obtain the function of
width and highness $\rho (t)$ whose time evolution is plotted as
Fig. 1. In the considered case the module oscillates with very
small amplitude that implies width and highness of the wave-packet
train being changed also quite small.
\begin{figure}[htbp]
\centering
\includegraphics[width=2.5in]{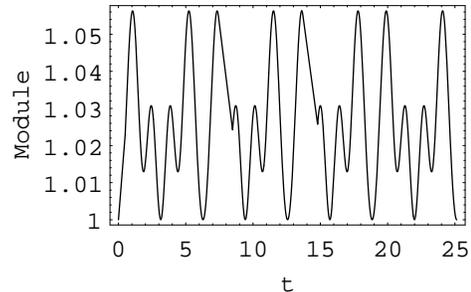}
\vspace{-0.5cm} \caption{Time evolution of the module $\rho(t)$ of
the classical solution that exhibits small change of the solitonic
width and highness with period $2\pi$. The space-time variables
$x_c$ and $t$ are normalized in units $l_h$ and $\omega^{-1}$
respectively.}\label{fig1}
\end{figure}

Substituting Eq. (14) into Eq. (7) and further into Eq. (12) leads
to a determined form of the module $R_n(x, t)$. Using this form we
numerically plot the probability density $R_n^2$ versus $x$ for
$n=8$ at the times $t=0, \ \pi/2, \ 2\pi$ respectively as Fig. 2a,
2b, and 2c. These plots show that the solitonlike train propagates
from $x_c=-10$ to $x_c=10$ and approximately keeps its shape. In
the next half period, $2\pi \le t \le 4\pi$, an inverse process
will occur.
\begin{figure*}[htbp]
\centering
\includegraphics[width=2.2in]{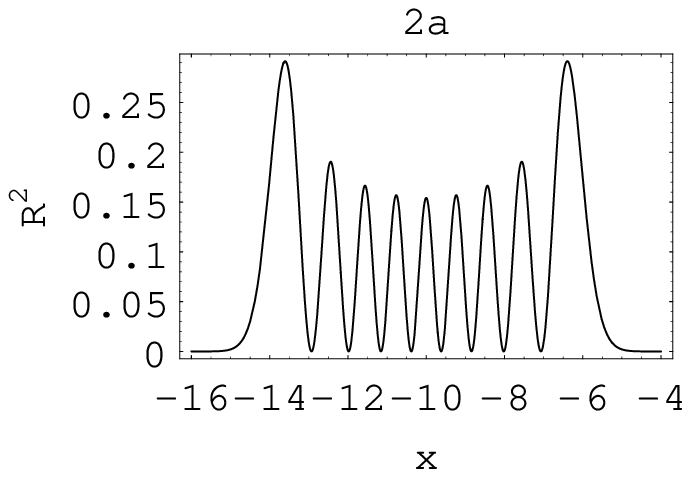}
\includegraphics[width=2.2in]{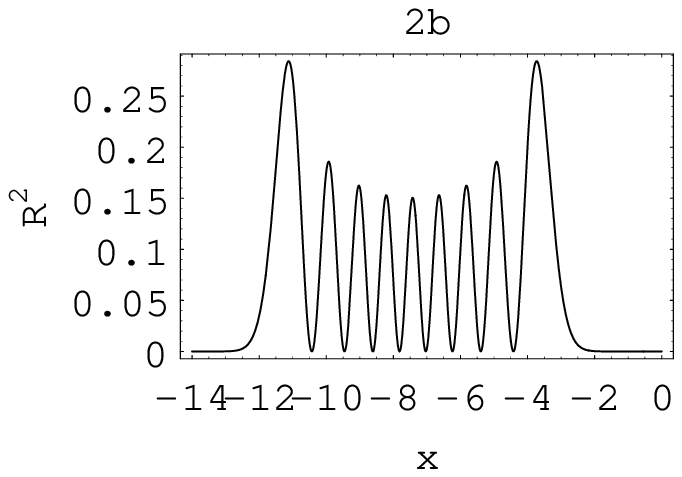}
\includegraphics[width=2.2in]{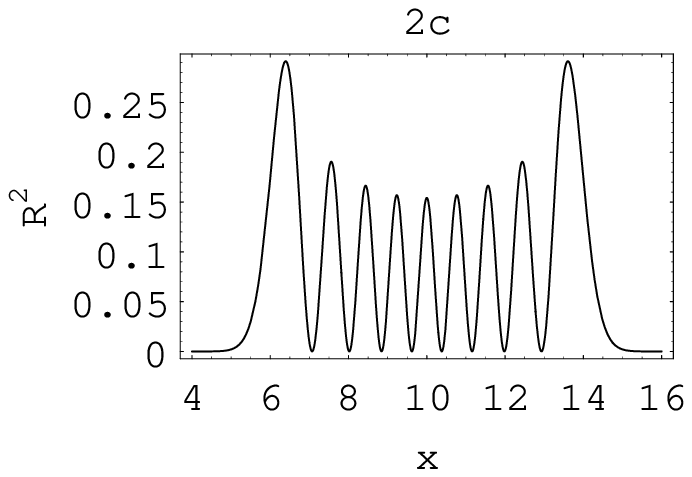}
\caption{The probability density of solitonlike train $R_8^2$
versus $x$ from Eq. (12) for (a) $t=0$, (b) $t=\pi/2$ and (c)
$t=2\pi$. Nine wave-packets are shown in propagation with small
change of shape. The space-time variables $x$ and $t$ are
normalized in units $l_h$ and $\omega^{-1}$, and the probability
density is normalized by $l_h^{-1}$.}\label{fig2}
\end{figure*}

\bf Collapse and revival of the wave-packets \rm  From Eq. (13) we
know that when small constants $A$ and $b_0/c_0$ are applied to
Eq. (12), the wave-packet trains will approximately fix their
centers. Further we let constant $B$ be much greater than $A$,
Eqs. (6) and (7) could lead to large change of the function
$\rho(t)$ for $\alpha - \beta =\pi/2$. This will cause
consequently large changes of the widthes and highnesses of the
wave-packet trains described by Eq. (12). The large deformations
can be identified with the so-called collapse and revival of the
wave-packets \cite{Zeng}. As an example, we take the parameter set
$\alpha =0, \ \beta =-\pi/2, \ A=b_0=0.02, \ B=10, \ c_0=1$, from
Eqs. (6), (7) and (13) to solve for $\rho (t)$ and $x_c(t)$. The
center of the wave-packet train could oscillate with amplitude
only in $Ab_0=10^{-4}$ order. The function of width and highness
$\rho (t)$ periodically changes with amplitude 10 and minimum
value 0.02. The large amplitude and small minimum value of $\rho
(t)$ mean periodical collapse and revival of the wave-packet
train. Using these parameters, from Eq. (12) we plot spatial
evolution of the probability density with $n=4$ for several
different times as in Fig 3. This figure displays that the
initially higher and narrower wave-packet train is collapsed to
very low and width at $t=\pi$ and revived to initial state at
$t=2\pi$, as in Fig. 3b and 3c respectively.
\begin{figure*}[htbp]
\centering
\includegraphics[width=2.2in]{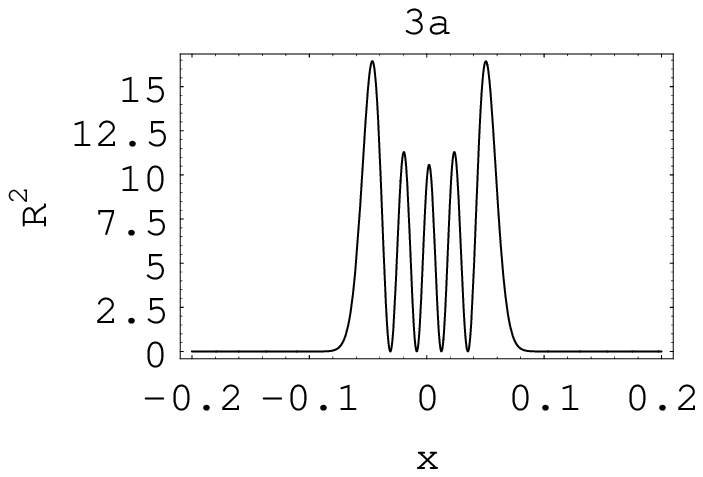}
\includegraphics[width=2.2in]{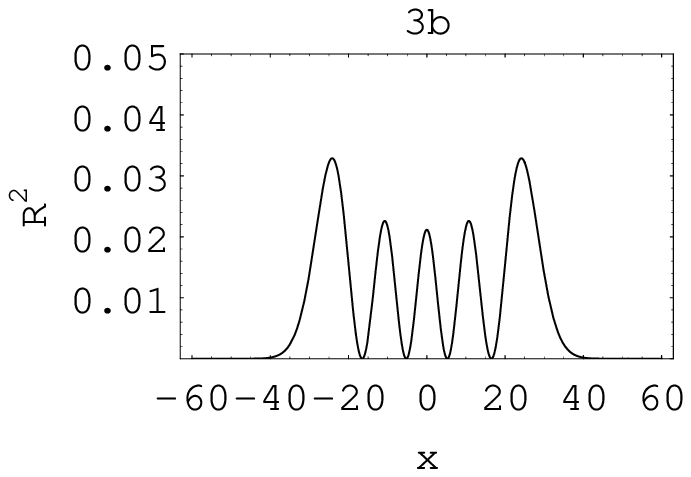}
\includegraphics[width=2.2in]{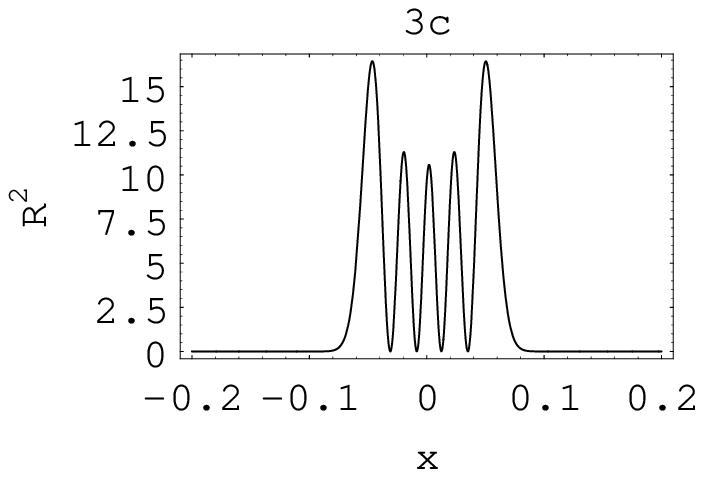}
\caption{The collapse and revival of wave-packet train from Eq.
(12) with $n=4$ for (a) $t=0$, (b) $t=\pi$, (c) $t=2\pi$. The
wave-packet train consisting five packets is shown with
approximately fixed center and great deformation. The space-time
coordinates and probability density are normalized in the same
units with Fig. 2.}\label{fig3}
\end{figure*}

In conclusion, we have investigated the quantum and classical
motions of a time-dependent harmonic oscillator, which is
associated with a Paul-trapped ion system. A new kind of exact
solutions of the quantum-mechanical Schr$\ddot o$dinger equation
is constructed by using the corresponding exact solutions of the
classical-mechanical Mathieu equation, which describes
propagations and deformations of the wave-packet trains. When the
deformations are small, they behave like some soliton trains. The
large deformations are identified with collapse and revival of the
wave-packet trains. If we select the initial conditions to fix
center of the wave-packet trains, the result is reduced to Brown's
quantum motion \cite{Brown}. Our ground state solution with $n=0$
seems to be Cessa's coherent state \cite{Cessa}. The state
$\psi_1$ of two wave-packets is similar to the Schr$\ddot
o$dinger's cat state \cite{Monroe}. We desire the new exact
solutions to play an important role in treating various
harmonically confined systems, e.g. the Bose-Einstein condensate
held in a magnetic well.

\bf Acknowledgement \rm  This work was supported by the NNSF of
China under Grant No. 10275023 and the NLMRAMP of China under
Grant No. T152103, and by the Hubei Provincial Key Laboratory of
Gravitation and Quantum Physics of China.

\end{document}